\documentclass{article}
\usepackage{naturetex}
\usepackage{amsthm}

\newtheorem{lemma}{Lemma}

\begin{document}

\maketitle

{\setstretch{1.0}
	\section*{Abstract}
	Precise	quantum key distribution (QKD) secure bound analysis is essential for practical QKD systems. The effect of uniformity of random number seed for privacy amplification is not considered in existing secure bound analysis. In this paper, we propose and prove the quantum leftover hash lemma with non-uniform random number seeds based on the min-entropy, and we give a precise QKD secure bound analysis with non-uniform random number seeds on this basis. We take the two-decoy BB84 protocol as an example to simulate the effect of random number seed uniformity on the secure bound of a QKD system. The experimental results indicate that when the average min-entropy of the random number generator is below 0.95, the secure bound of a QKD system will be seriously affected.
	
}

\newpage
\section*{Introduction}

Quantum key distribution (QKD) technology provides secure communication service with information-theoretic security~\cite{BennettCharlesandBrassard1984}. As the development of QKD technology, QKD has moved towards the practical stage. The pracical security of QKD systems has gradually attracted researcher's attention, and many ideal assumptions in QKD security analysis are found unsatisfied in practical QKD systems\cite{Tomamichel2012,Gottesman2004,2016Decoy}. One of these assumptions is that the random number seeds used for privacy amplification in a QKD system must be strictly uniformly distributed, and this is very difficult to guarantee in an actual system\cite{Tomamichel2011}. This gap may seriously affect the security of privacy amplification, which in turn seriously affects the secure bound of QKD. However, the exact extent of this impact has not been analyzed.

Privacy amplification is a necessary part of a QKD system. It is the art of distilling a information-theoretic secure key from a partially
secure string with a hash function by public discussion between two parties~\cite{Bennett1995}. In order to ensure the security of keys, the hash function must be randomly selected from a universal hash family with random number seeds in the existing PA secure proof~\cite{Tomamichel2011}. Hayashi et al. quantifies the uniformity of random number seeds with min-entropy, and analyzes the effect of min-entropy of random number seeds  on privacy amplification security under classical information theory\cite{Hayashi2016}. However, there is still a lack of security analysis under quantum information theory and analysis of the impact of random seed min-entropy on secure bound of QKD.

Aiming at this problem, this paper proposes and proves the quantum leftover hash lemma with non-uniform random number seeds, and analyzes a precise QKD secure bound with non-uniform random number seeds. 

In order to further analyze the influence of PA random number seeds on the secure key rate of QKD systems, we investigate the average min-entropy of random number generators in existing QKD systems. We find that most systems do not give the average minimum entropy of their random seeds. Therefore, we investigated and tested the min-entropy of some commonly used random number generators in QKD systems. We found that these random number generators could not achieve the perfect minimum entropy, so they would have a obvious impact on the secure key rate.   

\newpage
\section*{Results}\label{results}
\subsection*{Quantum leftover hash lemma with non-uniform random seeds}

We discussed the security of QKD under quantum information theory and universal composable security. The security of a QKD protocol should be considered on the secrecy and correctness.

Suppose the information possessed by the eavesdropper is $E$, then a key relative to the eavesdropping information $E$ can be called ${\epsilon}_{sec}-$ secrecy, when the statistical distance between the key and a key that is uniformly distributed and independent of $E$ is less than ${\epsilon}_{sec}$:
\begin{equation}
	\frac{1}{2}{\left\| {{\rho _{{\rm{SE}}}} - {S_U} \otimes {\rho _{E}}} \right\|_1} \le {\epsilon_{sec}}.
\end{equation} 

In universal composable security theory, key correctness represents the probability that $S_A$ and $S_B$ are different:
\begin{equation}
{Pr}[S_A \ne S_B] \le \epsilon_{cor}
\end{equation} 

Considering both secrecy and correctness, when the key is $\epsilon_{sec}$-secrecy and $\epsilon_{cor}$-correctness, the key is $\overline \epsilon-$secure:

\begin{equation}
	\overline \epsilon = \epsilon_{sec} + \epsilon_{cor}.
\end{equation} 

We proposed and proved the quantum leftover hash lemma with non-uniform random seeds under quantum information theory and universal composable security:

\textbf{Theorem 1 (Quantum Leftover Hash Lemma With Non-Uniform Random Seeds)} Let $F_R$ be a universal hashing family of functions from $X$ to $S$, $f_r$ is a hash function randomly selected from $F_R$ with random seeds $R \in \{0,1\}^\alpha$, $|F_R|=2^{\alpha}$ and $P_{F_R}$ satisfies $H_{min}(P_{F_R}) \ge \beta$, and $s=f_r(x)$. Let ${\rho _{XE}} = \sum\limits_x {\left| x \right\rangle {{\left\langle x \right|}_X} \otimes \rho _E^{[x]}} $  
and cq-states ${\rho _{{F_R}{\rm{S}}E}} = \sum\limits_{{f_r}} {\sum\limits_{\rm{s}} {{P_{{F_R}}}\left| {{f_r}} \right\rangle \langle {f_r}{|_{{F_R}}} \otimes } } \left| s \right\rangle \langle s{|_S} \otimes \rho _E^{\left[ {{f_r},s} \right]}$. Then for any $\epsilon  \ge 0$,
\begin{equation}
\Delta  = \sum\limits_{{f_r}} {{P_{{F_R}}}({f_r}){D_u}{{(S|E)}_{{\rho ^{[{f_r}]}}}}}\le\frac{1}{2} \times {2^{\alpha  - \beta }} \times {2^{ - \frac{1}{2}(H_{\min }^\varepsilon ({\rho _{{\rm{XE}}}}\left| E \right.) - l)}} + \varepsilon,
\end{equation}
where $E$ is the side information of eavesdropper.

More importantly, we further analyzed the effect of random number seed uniformity on the secure bound of a QKD protocol, and the secure bound of a QKD system with non-uniform random number seed is obtained as follow,
\begin{equation}
l \le H_{\min }^\varepsilon ({\rho _{SE'}}\left| {E'} \right.) - lea{k_{EC}} - 2{\log _2}\frac{1}{{2(\varepsilon_{sec}  - \varepsilon )}}-\log_{2}{\frac{2}{\epsilon_{cor}}}-(\alpha-\beta).
\end{equation}

For further analyzing the influence of PA random number seeds on the secure key rate of QKD systems, we investigated and tested the min-entropy of some commonly used random number generators in QKD systems as shown in Table \ref{tab:1}.

\begin{table}
	\centering
	\caption{The average min-entropy of common random number generator}
	\label{tab:1}       
		\begin{tabular}{ccccc}
			\hline
			\hline
			Random Number Generator  & Type & Refer/Test & Test Scale & Average Min-entropy \\
			\hline
			\hline
			IDQ Quantis-PCIe-40M  & QRNG & Test & 100Mb  & 0.990 \\
			MATLAB unifrnd        & PRNG & Test & 100Mb  & 0.988 \\
			Random.org       	  & TRNG & Refer & --  & 0.931 \\
			Intel DRNG		      & TRNG & Refer & --  & 0.930 \\
			\hline
			\hline
		\end{tabular}        
\end{table}

\begin{figure}[h]
	\begin{center}
		\includegraphics[width=0.9\textwidth]{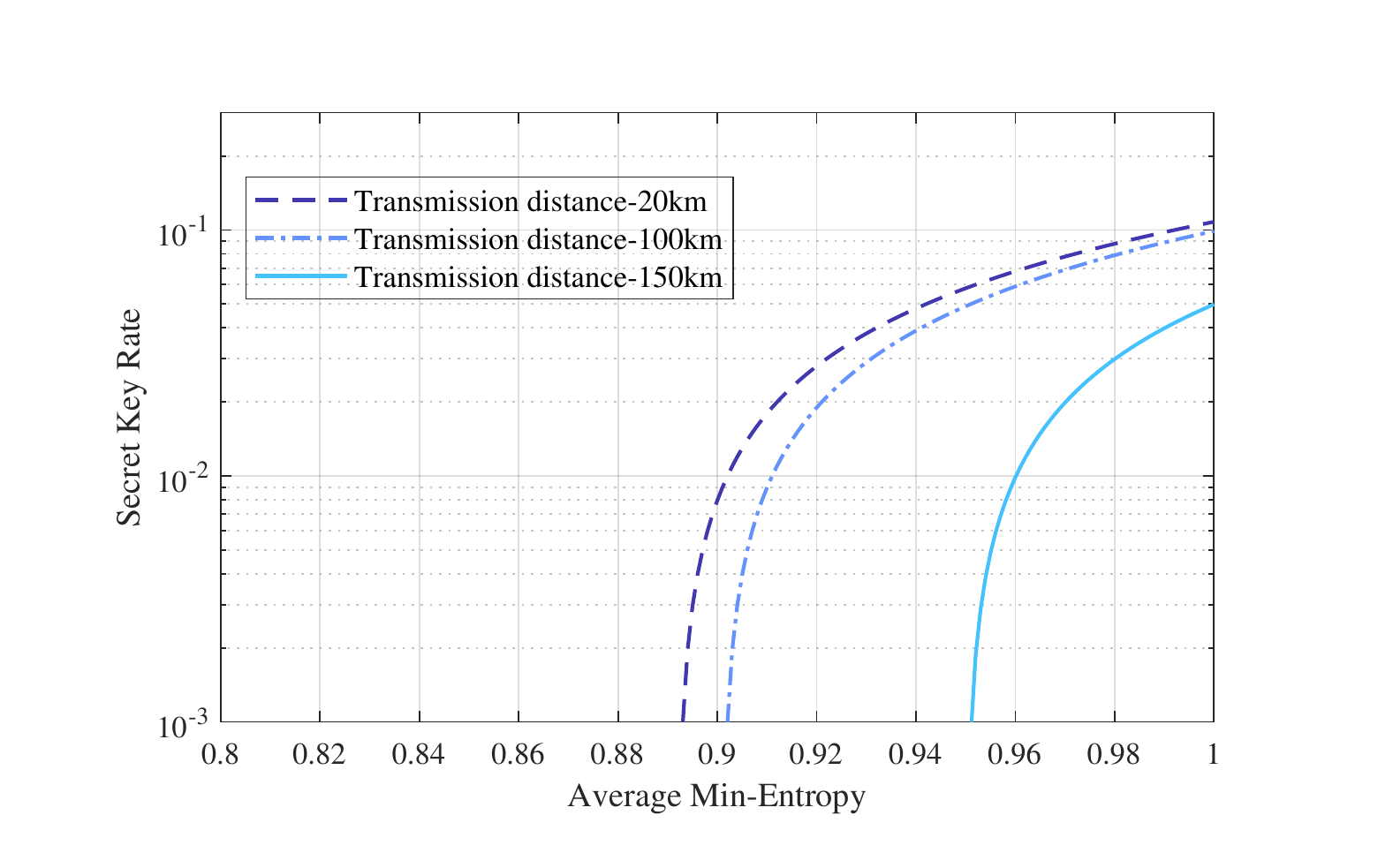}
	\end{center}
	\caption{The relation between random uniformity and SKR under different distances}
	\label{fig:my_plot1}
\end{figure}

\begin{figure}[h]
	\begin{center}
		\includegraphics[width=0.9\textwidth]{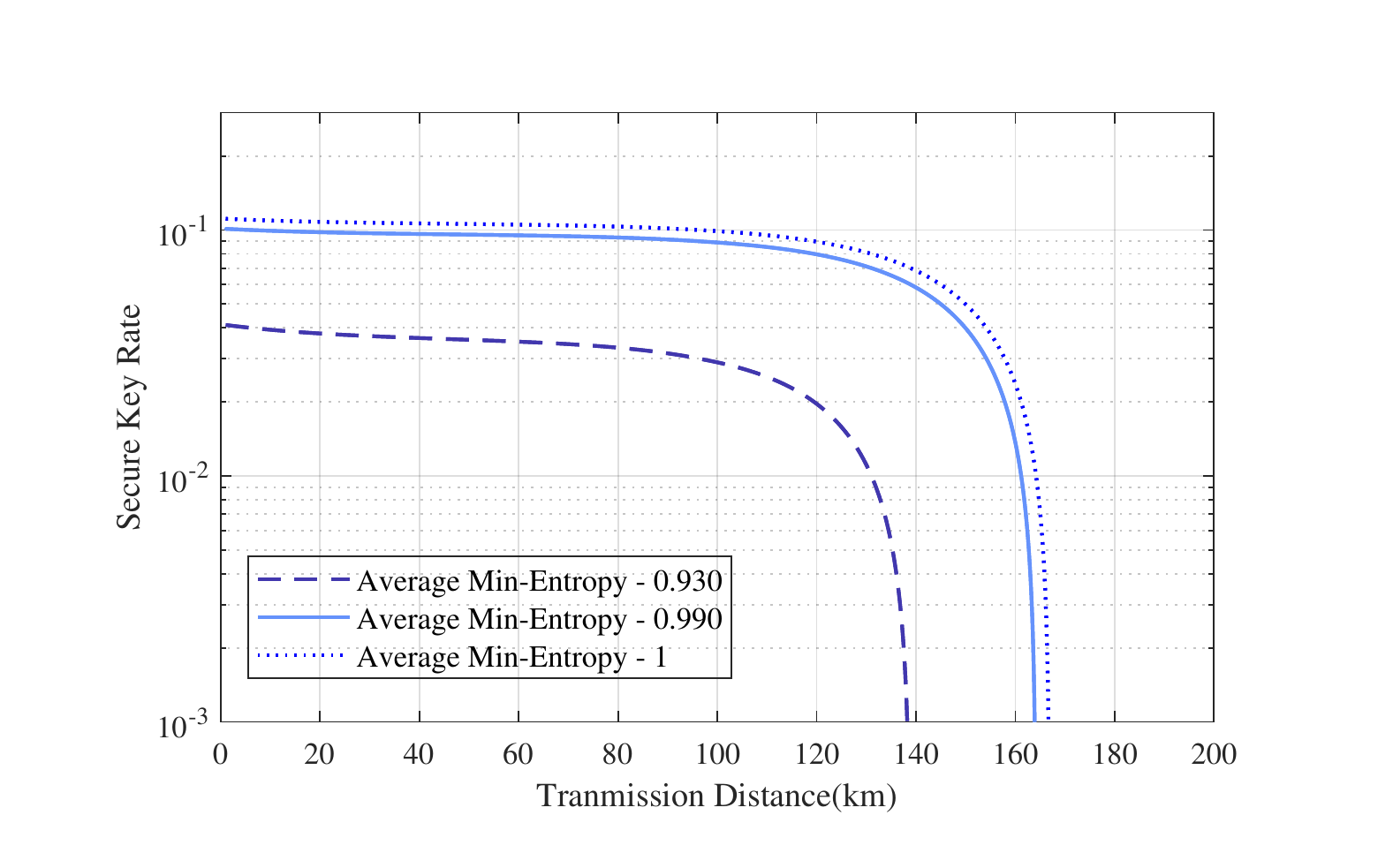}
	\end{center}
	\caption{The relation between random uniformity and SKR under different distances}
	\label{fig:my_plot2}
\end{figure}

We refer to a typical decoy BB84 protocol to experiment the effect of random number min-entropy on the QKD secure key rate. The experiment result is indicated as Fig. \ref{fig:my_plot1} and Fig. \ref{fig:my_plot2}.

The above experimental results indicate that, (1) the average min-entropy of the random number generator is below 0.95, the secure bound of a QKD system will be seriously affected; (2) Most commonly used random number generators in a QKD system will influence the secret key rate of QKD seriously.


\newpage
\section*{Methods}

The proof of quantum leftover hash lemma with non-uniform random seeds is given as below.

\textbf{Theorem 1 (Quantum Leftover Hash Lemma With Non-Uniform Random Seeds)} Let $F_R$ be a universal hashing family of functions from $X$ to $S$, $f_r$ is a hash function randomly selected from $F_R$ with random seeds $R \in \{0,1\}^\alpha$, $|F_R|=2^{\alpha}$ and $P_{F_R}$ satisfies $H_{min}(P_{F_R}) \ge \beta$, and $s=f_r(x)$. Let ${\rho _{XE}} = \sum\limits_x {\left| x \right\rangle {{\left\langle x \right|}_X} \otimes \rho _E^{[x]}} $  
and cq-states ${\rho _{{F_R}{\rm{S}}E}} = \sum\limits_{{f_r}} {\sum\limits_{\rm{s}} {{P_{{F_R}}}\left| {{f_r}} \right\rangle \langle {f_r}{|_{{F_R}}} \otimes } } \left| s \right\rangle \langle s{|_S} \otimes \rho _E^{\left[ {{f_r},s} \right]}$. Then for any $\epsilon  \ge 0$,
\begin{equation}
\Delta  = \sum\limits_{{f_r}} {{P_{{F_R}}}({f_r}){D_u}{{(S|E)}_{{\rho ^{[{f_r}]}}}}}\le\frac{1}{2} \times {2^{\alpha  - \beta }} \times {2^{ - \frac{1}{2}(H_{\min }^\varepsilon ({\rho _{{\rm{XE}}}}\left| E \right.) - l)}} + \varepsilon,
\end{equation}
where $E$ is the side information of eavesdropper.

\begin{proof}

For, 
\begin{equation}
	\Delta  = \sum\limits_{{f_r}} {{P_{{F_R}}}({f_r}){D_u}{{(S|E)}_{{\rho ^{[{f_r}]}}}}},
\end{equation}
As $P_{F_R}$ satisfies $H_{min}(P_{F_R}) \ge \beta$, then for any ${{P_{{F_R}}}({f_r})}$, it satisfies ${{P_{{F_R}}}({f_r})} \le 2^{-\beta}$, then,
\begin{equation}
\Delta  = \sum\limits_{{f_r}} {{P_{{F_R}}}({f_r}){D_u}{{(S|E)}_{{\rho ^{[{f_r}]}}}}}\le \sum\limits_{{f_r}} {{2^{ - \beta }}{D_u}{{(S|E)}_{{\rho ^{[{f_r}]}}}}}
= {2^{\alpha  - \beta }}\sum\limits_{{f_{\rm{r}}}} {{2^{ - \alpha }}{D_u}{{(S|E)}_{{\rho ^{[{f_r}]}}}}},
\end{equation}

Since the set sizes of $F_R$ and $F_U$ are the same as $2^\alpha$, and the uniform distribution of $F_U$ satisfies $P_{F_u}(f_u)=2^{-\alpha}$, it can be obtained:
\begin{equation}
\Delta = \sum\limits_{{f_r}} {{P_{{F_R}}}({f_r}){D_u}{{(S|E)}_{{\rho ^{[{f_r}]}}}}}\le {2^{\alpha  - \beta }} \sum\limits_{{f_{\rm{u}}}} {{P_{{F_u}}}({f_u}){D_u}{{(S|E)}_{{\rho ^{[{f_u}]}}}}}
= {2^{\alpha  - \beta }}{D_u}{(S|{F_u}E)_\rho }.
\end{equation}
Further, according to Lemma 1, the upper limit of $\Delta$ can be obtained as:
\begin{equation}
\Delta = \sum\limits_{{f_r}} {{P_{{F_R}}}({f_r}){D_u}{{(S|E)}_{{\rho ^{[{f_r}]}}}}} \le \frac{1}{2} \times {2^{\alpha  - \beta }} \times {2^{ - \frac{1}{2}(H_{\min }^\varepsilon ({\rho _{{\rm{XE}}}}\left| E \right.) - l)}} + \varepsilon
\end{equation}
\end{proof}

In the above proof, this paper adopts the method of directly scaling ${{D_u}{{(S|F_{u}E)}_\rho }}$ to find its upper limit. Another more intuitive way is to directly scale the maximum collision probability of the approximate general hash to find the upper limit. The specific process is as follows.

First, according to the following lemma, the upper limit of ${{D_u}{{(S|F_{u}E)}_\rho }}$ can be obtained.

\begin{lemma}
	Let ${\rho _{AB}} \in {S_ \le }({{\bf{{\rm H}}}_{AB}})$, ${\tau _B} \in {S_ \le }({{\bf{{\rm H}}}_B})$ and $\sup \{ {\tau _B}\}  \supseteq \sup \{ {\rho _B}\} $, then, 
	\begin{equation}
	{{D_u}{{(S|FE)}_\rho }} \le \frac{1}{2}\sqrt {{d_A}{\Gamma _C}({\rho _{AB}}|{\tau _B}) - tr({\rho _B}\tau _B^{ - 1/2}{\rho _B}\tau _B^{ - 1/2})},
	\end{equation}
	where $d_A$ is the set size of $A$.
\end{lemma} 

According to Lemma 1, the upper limit of ${{D_u}{{(S|F_{R}E)}_\rho }}$ can be obtained,
\begin{equation}
	{{D_u}{{(S|F_{R}E)}_\rho }} \le \frac{1}{2}\sqrt {{2^l}{\Gamma _C}({\rho _{FSE}}|{\rho _F} \otimes {\tau _E}) - tr({\rho _E}\tau _E^{ - 1/2}{\rho _E}\tau _E^{ - 1/2})}.
\end{equation}

Then, by scaling ${{\Gamma _C}({\rho _{FSE}}|{\rho _F} \otimes {\tau _E})}$ to find its upper limit, it can get,

\begin{equation}
	\begin{array}{*{20}{l}}
	{{\Gamma _C}({\rho _{{F_R}SE}}|{\rho _F} \otimes {\tau _E})}\\
	{ = \sum\limits_{f \in {F_R}} {{P_{{F_R}}}} \sum\limits_z {tr\left( {\left| {{f_r}} \right\rangle \langle {f_r}{|_{{F_R}}} \otimes \left| s \right\rangle \langle s{|_S} \otimes \rho _E^{\left[ {{f_r},s} \right]}\tau _E^{ - 1/2}\rho _E^{\left[ {{f_r},s} \right]}\tau _E^{ - 1/2}} \right)} }\\
	{ = \mathop {\rm{E}}\limits_{{f_r} \in {F_R}} \left[ {\sum\limits_z {tr\left( {\rho _E^{\left[ {{f_r},s} \right]}\tau _E^{ - 1/2}\rho _E^{\left[ {{f_r},s} \right]}\tau _E^{ - 1/2}} \right)} } \right]}\\
	{ = \sum\limits_{x,x'} {\mathop {\rm{E}}\limits_{{f_r} \in {F_R}} \left[ {\sum\limits_z {{\delta _{{f_r}(x) = z}}{\delta _{{f_r}(x') = z}}} } \right]} tr\left( {\rho _E^{\left[ x \right]}\tau _E^{ - 1/2}\rho _E^{\left[ {x'} \right]}\tau _E^{ - 1/2}} \right).}
	\end{array}
\end{equation}

According to the definition of the $\delta$-almost universal family, when the random number seed satisfies the uniform distribution, the above expectation satisfies $\mathop {\rm E}\limits_{f \in {F_u}} \left[ { \sum\limits_z {{\delta _{f(x) = z}}{\delta _{f(x') = z}}} } \right] \le \delta $. When the random number seed does not satisfy the uniform distribution, it can be scaled to get:
\begin{equation}
	\begin{array}{*{20}{l}}
	{\mathop {\rm{E}}\limits_{{f_r} \in {F_R}} \left[ {\sum\limits_z {{\delta _{{f_r}(x) = z}}{\delta _{{f_r}(x') = z}}} } \right]}\\
	{ = \sum\limits_{{f_r} \in {F_R}} {{P_{{F_R}}}} \left[ {\sum\limits_z {{\delta _{{f_r}(x) = z}}{\delta _{{f_r}(x') = z}}} } \right]}\\
	{ \le \sum\limits_{{f_r} \in {F_R}} {{2^{ - \beta }}} \left[ {\sum\limits_z {{\delta _{{f_r}(x) = z}}{\delta _{{f_r}(x') = z}}} } \right]}\\
	{ = {2^{\alpha  - \beta }}\sum\limits_{{f_r}} {{2^{ - \alpha }}} \left[ {\sum\limits_z {{\delta _{{f_r}(x) = z}}{\delta _{{f_r}(x') = z}}} } \right]}\\
	{ \le {2^{\alpha  - \beta }} \times \delta }
	\end{array}
\end{equation}

According to this result, the upper limit of ${\Gamma _C}({\rho _{{F_R}SE}}|{\rho _{{F_R}}} \otimes {\tau _E})$ is,
\begin{equation}
	{\Gamma _C}({\rho _{{F_R}SE}}|{\rho _{{F_R}}} \otimes {\tau _E}) \le {\Gamma _C}({\rho _{XE}}\left| {{\tau _E}} \right.) + {2^{\alpha  - \beta }} \times \delta  \times {\rm{tr}}\left( {{\rho _E}\tau _E^{ - 1/2}{\rho _E}\tau _E^{ - 1/2}} \right).
\end{equation}

Let ${\rho _E} = {\tau _E}$, the formula can be further simplified as:
\begin{equation}
	{\Gamma _C}({\rho _{{F_R}SE}}|{\rho _{{F_R}}} \otimes {\tau _E}) \le {\Gamma _C}({\rho _{XE}}\left| {{\tau _E}} \right.) + {2^{\alpha  - \beta }} \times \delta  \times {\rm{tr}}{\rho _E}.
\end{equation}

Then,
\begin{equation}
	{{D_u}{{(S|F_{R}E)}_\rho }} \le \frac{1}{2} \sqrt {{2^l}{\Gamma _C}({\rho _{XE}}\left| {{\rho _E}} \right.) + \left( {{2^{\alpha  - \beta }} \times \delta  \times {2^l} - 1} \right){\rm{tr}}{\rho _E}} .
\end{equation}

Substitute the smoothed minimum entropy, $\delta=2^{-l}$ and $\rm{tr}{\rho_{E}}$, we can get,
\begin{equation}
\Delta = \sum\limits_{{f_r}} {{P_{{F_R}}}({f_r}){D_u}{{(S|E)}_{{\rho ^{[{f_r}]}}}}}\le\frac{1}{2} \sqrt {{2^{l - {H_{{{\min }^\varepsilon }}}({\rho _{{\rm{XE}}}}\left| E \right.)}} + {2^{\alpha  - \beta }} - 1}  + \varepsilon
\end{equation}

Comparing this upper limit with the upper limit in the proof, it can be found that this upper limit is much higher than the upper limit in the proof, indicating that although the scaling idea of this method is more obvious, the scaling method in the proof in this paper obtains a tighter upper limit.

\newpage
\bibliography{bibliography}
\bibliographystyle{naturemag}

\end{document}